\documentclass{mn2e}
\usepackage{graphicx}

\newcommand{\beq}{\begin{equation}}
\newcommand{\eeq}{\end{equation}}

\begin{document}

\title{The shapes of molecular cloud cores in Orion}
%
\author[K. Tassis]
{Konstantinos Tassis$^{1,2}$\\
$^{1}$Department of Astronomy and Astrophysics, 
The University of Chicago, Chicago, IL 60637, USA\\
$^{2}$The Kavli Institute for Cosmological Physics, 
The University of Chicago, Chicago, IL 60637, USA \\
}
\maketitle

\label{firstpage}
\begin{abstract}

We investigate the intrinsic shapes of starless cores in the Orion GMC, using
the prestellar core sample of Nutter \& Ward-Thompson (2007), which is based on 
submillimeter SCUBA data. We employ a
maximum-likelihood method to reconstruct the intrinsic distribution of
ellipsoid axial ratios from observations of the axial ratios of
projected plane-of-the-sky core ellipses. We find that, independently
of the details of the assumed functional form of the distribution,
there is a strong preference for oblate cores of finite
thickness. Cores with varying finite degrees of triaxiality are a
better fit than purely axisymmetric cores although cores close to
axisymmetry are not excluded by the data. The incidence of prolate
starless cores in Orion is found to be very infrequent. 
We also test the consistency of the observed data with a uniform
distribution of intrinsic shapes, where oblate and prolate cores of
all degrees of triaxiality occur with equal probability. Such a
distribution is excluded at the $0.1\%$ level. These findings have
important implications for theories of core formation within molecular
clouds. 
\end{abstract}


\section{Introduction}

The intrinsic shapes of molecular cloud cores encode important clues
about the processes of molecular cloud fragmentation and core
formation. If cores form as the result of converging turbulent
flows (e.g., see review by MacLow \& Klessen 2004) then they
are expected to have random, triaxial shapes, with a slight preference for
prolateness (Gammie et al. 2003; Li et al. 2004). If, on the other hand,
magnetic fields are dynamically important, and ambipolar diffusion
mediates the core formation process, then the cores are expected to be
flattened along the magnetic field due to the extra magnetic
force acting perpendicular to the field lines. Core shapes in this
case are oblate (Mouschovias 1976) although not necessarily
axisymmetric (Basu \& Ciolek 2004; Ciolek \& Basu 2006). 
Unfortunately, the distribution of core shapes 
cannot be easily derived from observations,  
because it is not possible to directly deproject the
shape  of each core - statistical techniques have 
to be used instead.  
Early work on core shapes, assuming axial symmetry, seemed to favour prolate cores (Myers et al. 1991; Ryden 1996). However, subsequent investigations which relaxed the axisymmetry assumption have consistently yielded triaxial, preferentially oblate core shapes (e.g. Jones et al. 2001; Jones and Basu 2002; Goodwin et al. 2002), 
independently of tracer or core sample.

In light of the recent debate concerning the core formation timescale
and its use as a test for theories of core
formation in molecular clouds (Hartmann et al. 2001; Tassis \& Mouschovias 2004;
Ballesteros-Paredes \& Hartmann 2006; Mouschovias, Tassis \& Kunz
2006), the timing is especially opportune to revisit the question of
the intrinsic core shapes, as these may serve as an independent test
of theories of core formation. The recent progress in studies
of turbulence-driven core formation and the convergence of the
findings of different groups concerning the expected distribution of
core shapes in this scenario (Gammie et al. 2003; Li et al. 2004) also
offers an unprecedented opportunity for direct comparison between
theory and observations.  We approach the problem using a new
statistical method: a maximum-likelihood
analysis, aimed to {\em reconstruct} the intrinsic shape distribution 
of molecular cloud cores. Maximum-likelihood analyses are particularly
powerful because they allow treatment of different functional forms
of the intrinsic shape distribution, distinguishing between functional
forms, and explicitly account for observational uncertainties and
different assumptions about potential biases in viewing
angles. Although here we present a first treatment under the simplest
possible assumptions, the expansion of the formalism to include more
complicated scenarios is straightforward and we will pursue it in a
future publication. 

We use the recently released core dataset of Nutter \& Ward-Thompson
(2007) (NWT07), which offers the opportunity to work with a sample with a
series of unique features:  a high level of completeness 
(NWT07 use this sample to derive the
core mass function in the mass range $0.3$ to $\sim 100 {\rm
  M_\odot}$); cores of varying sizes and masses
identified in a single survey [in this case, a SCUBA ({\it
  Submillimeter Common User Bolometer Array}) survey of the Orion
Giant Molecular Cloud (GMC)]; confident discrimination between
prestellar and protostellar cores (protostellar cores were identified
by the authors using {\it Spitzer} data, and we have excluded such
cores from our analysis). The sample used here consists 
of the 286 prestellar cores from the Orion A
North and South and Orion B North and South 
star forming regions. Axial ratios for these objects are calculated
from the  quoted semi-major and semi-minor dimensions in NWT07. 

\section{Formalism}\label{form}

Consider a system of coordinates centered on a triaxial ellipsoid
model molecular cloud core with  principal axes ($a$, $b$, $c$). The
triaxial ellipsoid surface in this system obeys 
\beq\label{ONE}
\frac{x^2}{a^2} + \frac{y^2}{b^2} + \frac{z^2}{c^2} = 1\,.
\eeq
The image of the core appears on the
plane of the sky as an ellipse,  the properties of which
 can be calculated from Eq. (\ref{ONE})
and the orientation of the observer's line of sight
(Binney 1985). If $a \geq b \geq c$ and we define the axial 
ratios $\zeta \equiv b/a$ and $\xi \equiv c/a$, so $1\geq \zeta \geq \xi$, 
the plane-of-the-sky isophotes of the core 
will be coaxial ellipses of axial ratio
\begin{equation}\label{MOO}
q (\theta, \phi, \zeta, \xi) = 
\sqrt{\frac{A+C-\sqrt{(A-C)^2+B^2}}
{A+C+\sqrt{(A-C)^2+B^2}}} \, \leq 1
\end{equation}
where $\theta$ and $\phi$ are 
the line-of-sight orientation angles 
and 
\beq\label{THEA}
A \equiv \frac{\cos^2\theta}{\xi^2}
\left(\sin^2\phi +\frac{\cos^2\phi}{\zeta^2}\right)
+\frac{\sin^2\theta}{\zeta^2}\,,
\eeq

\beq\label{THEB}
B \equiv \cos \theta \sin 2\phi
\left(1-\frac{1}{\zeta^2}\right)\frac{1}{\xi^2}\,, \,
C \equiv \left(\frac{\sin^2\phi}{\zeta^2} + \cos^2 \phi
\right) \frac{1}{\xi^2}\,.
\eeq
The axial ratio $q$ has values limited between $0$ and $1$. A very
elongated ellipse has $q\ll 1$,
while a circle
has $q=1$. 

Let the cores of the Orion GMC have a single 
intrinsic distribution of shapes,  
$p_{\rm Orion}(\zeta,\xi) d\zeta d\xi$, defined as the 
fraction of cores with axial ratios between $\zeta$ and $\zeta
+d\zeta$ and $\xi$ and $\xi + d\xi$. 
Let us assume that
$p_{\rm Orion}$ can be parametrized by a set of 
parameters $C_i$\footnote{For example, in the case of a Gaussian, 
  $C_1$ and $C_2$ would be  the mean and the standard deviation respectively.}. 
From an assumed $p_{\rm Orion}(\zeta,\xi, C_i)$, we can 
calculate the distribution of
axial ratios of projected cloud ellipses,
$p(q)$, by assuming that the viewing
angles
are random ($\phi$ and 
$\cos \theta$ follow uniform distributions). For this calculation we
use the following Monte-Carlo method:

(i) We select a set of parameters $C_i$ for the distribution
      $p_{\rm Orion}(\zeta, \xi, C_i)$.

(ii)We randomly draw a pair of ($\zeta, \xi$) from the distribution
      $p_{\rm Orion}(\zeta, \xi, C_i)$, and 
      a pair of ($\theta$, $\phi$) from a uniform
      probability distribution (equal probability for any value of
      $\cos \theta$ or $\phi$ in the intervals $[-1,1]$ and $[0,2\pi]$
      respectively). We use the values of $\zeta, \xi, \theta, \phi$
      to calculate $q$ through Eq. (\ref{MOO})

(iii) We repeat steps (i)-(ii) a large number of times. We bin the obtained
      values of $q$, and we normalize the obtained distribution $p(q)$ so
      that it is a probability density function (pdf) ($\int_1^\infty p(q,
      C_i)dq = 1$). 
 
Once we have obtained $p(q, C_i)$, we can calculate the likelihood function
$\mathcal{L}$. If we have $n$ observations of core
axial ratios $q_1, q_2, ... q_n$ then, dropping constant normalization 
factors, the likelihood is
\begin{equation}
\mathcal{L} (C_i, q_1, q_2,  ...., q_n) = 
\prod_{j=1}^n p(q_j, C_i)
\end{equation} 
The maximum-likelihood parameters of the intrinsic shape distribution
$p(\zeta, \xi, C_i)$ are that set of $C_i$ which maximize the
likelihood $\mathcal{L}$ for our given set of observed axial ratios. 

Performing a likelihood
analysis  requires a selection of a functional form for the intrinsic
shape distribution $p(\zeta,\xi)$. Lacking any {\it a priori} knowledge on the functional form of $p(\zeta,\xi)$, we have chosen functional forms based on their simplicity and appropriateness of their properties (defined in finite domains, singly-peaked). We will perform this analysis
using two {\em distinct} families of distributions to test the sensitivity of
our result to the assumed functional form. 
In the following sections, we briefly discuss these functional forms.

\subsection{Beta Distribution of Axial Ratios}

Since both $\zeta$ and
$\xi$ are defined in finite domains, a natural choice is the beta
distribution, which is routinely used to model events which are constrained within finite value intervals. The probability density function of the beta family of distributions is $p(x) = x^{a-1}(1-x)^{b-1}/B(a,b)$, where
$B(a,b) = \int_0^1 t^{a-1}(1-t)^{b-1} dt$ is the beta function. The distribution is defined in the interval [0,1] and its shape is controlled by the non-negative shape parameters $a,b$. When $a,b>1$ the distribution is singly-peaked.

 We construct a joint distribution $p(\zeta, \xi)$ 
 assuming that $\zeta$ follows a beta distribution with domain 
$(0,1]$ 
and $\xi$ follows a beta distribution with 
domain $(0,\zeta]$.
We start from a double beta 
distribution in variables $x$ and $y$, which vary between $0$ and $1$,
with shape parameters ($a_x, b_x$) and  ($a_y, b_y$), respectively. 
With the change of 
variables $\zeta=x$ and $\xi = \zeta y$ 
[which maps the $(0,1]$ domain of $y$ to the desired 
$(0,\zeta]$ domain of $\xi$], we obtain the joint distribution for $\zeta,\xi$: 
\begin{eqnarray}\label{daughter}
p(\zeta,\xi) &=& 
\frac{\zeta^{a_x-a_y-b_y}
(1-\zeta)^{b_x-1} \xi^{a_y-1} (\zeta-\xi)^{b_y-1}
}{B(a_x,b_x)B(a_y,b_y)} \,.\end{eqnarray} 
The joint probability of 
Eq. (\ref{daughter}) is such that $\zeta$ and $\xi$ are {\em not} independent. 
The distribution of Eq. (\ref{daughter}) has a single peak at 
\begin{equation}\label{maximumzx}
\xi_{0} = \frac{a_y-1}{a_y+b_y-2}\zeta_0\,, \,\,\,\,\,\,\,\,\,\,
\zeta_{0}= \frac{C_1C_2-b_y+1}{C_1C_3-b_y+1}
\end{equation}
provided
that $a_y, b_y>1$ and that $a_x,b_y$ are such that $0<\zeta_0<1$. In 
Eq. (\ref{maximumzx}) $C_1 = (1-b_y)/(a_y+b_y-2)$, $C_2 =
a_x-a_y-b_y$, and $C_3 = a_x-a_y+b_x-b_y-1$. In
our analysis, we only admit values of the shape parameters that
result in a singly-peaked distribution. 
This is a tetra-parametric distribution and the likelihood analysis is
aimed at determining the values of $a_x$, $b_x$, $a_y$, and
$b_y$. 
\subsection{Modified Lognormal}

We construct this distribution by 
seeking an appropriate minimal transformation which transforms the
domain of $\zeta$ and $\xi$ from $(0,1)$ to $(-\infty,\infty)$;
we then take the transformed variables to follow a Gaussian distribution. 
Such an appropriate transformation is $\zeta\rightarrow x$ and $\xi
\rightarrow y$ where 
$x = \ln [\zeta/(1-\zeta)]$
and 
$y = \ln [\xi/(\zeta - \xi)]$.
Assuming a double Gaussian pdf of $x$ and $y$, 
the joint pdf of $\zeta$ and $\xi$ is
\begin{eqnarray}
p(\zeta,\xi) &=& \frac{
\exp\left[-\frac{(\ln\frac{\zeta}{1-\zeta}-x_0)^2}{2\sigma_x^2}\right]
\exp\left[-\frac{(\ln\frac{\zeta}{1-\zeta}-x_0)^2}{2\sigma_x^2}\right]}
{2\pi\sigma_x\sigma_y\xi(\zeta-\xi)(1-\zeta)}
\end{eqnarray}


This is also a tetra-parametric distribution and the likelihood analysis is
aimed at determining the values of $x_0$, $y_0$, $\sigma_x$, and
$\sigma_y$. Due to the transformation used to project the
$(0,1)$ interval to $(-\infty,\infty)$, the distribution in ($\zeta,\xi$)
space is skewed with respect to the original Gaussian shape of the
distribution in ($x,y$) space. The distortion increases as $\sigma_x$
and $\sigma_y$ increase, and eventually the distribution acquires
additional local maxima. To avoid
such extreme distortions
we limit the range of $\sigma_x$ and
$\sigma_y$ by imposing a prior requirement in our likelihood analysis
that $\sigma_x, \sigma_y \leq 0.7$. 

\section{Results}\label{theres}

\begin{figure}
\includegraphics[width=2.8in]{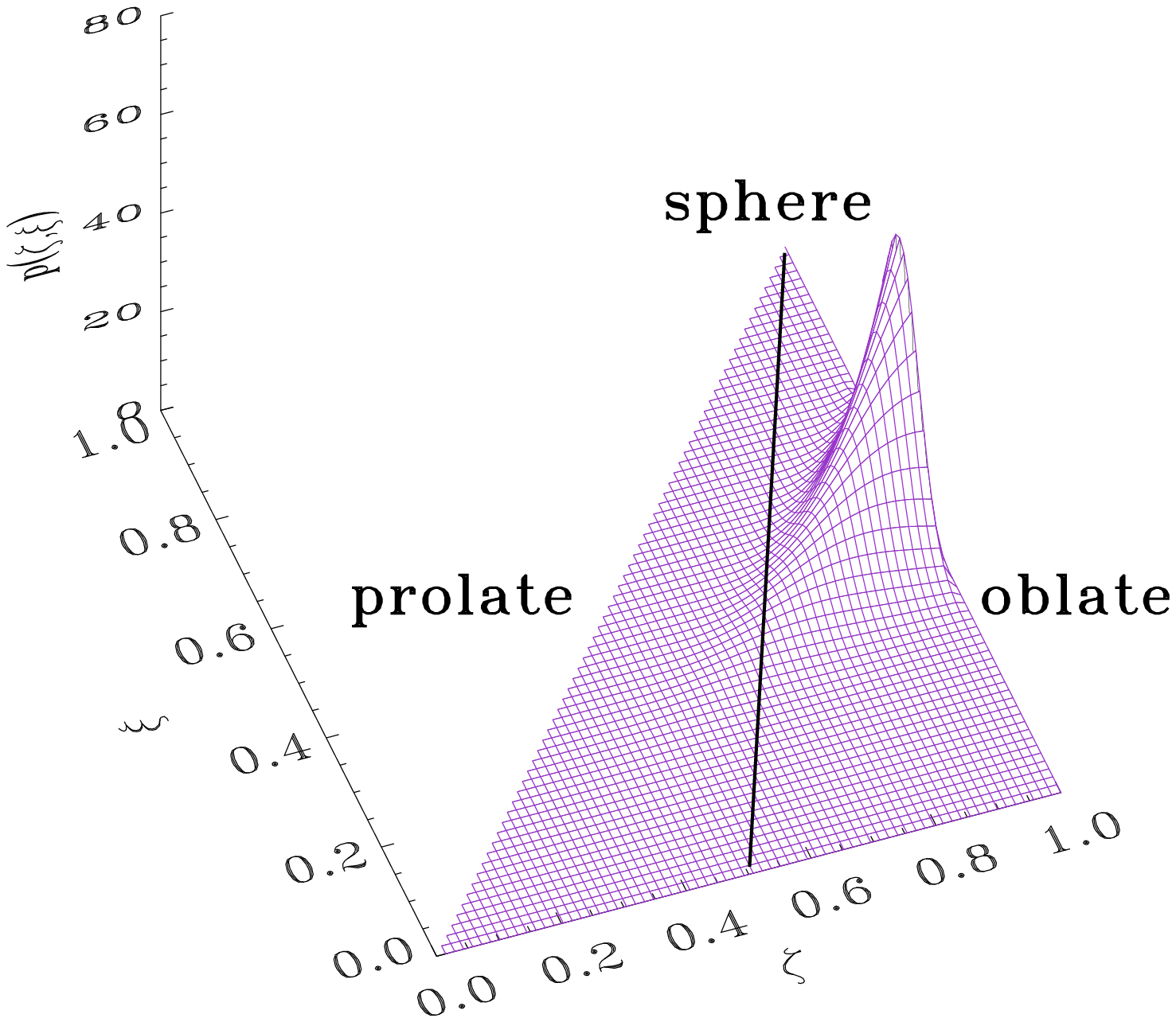}
\includegraphics[width=2.8in]{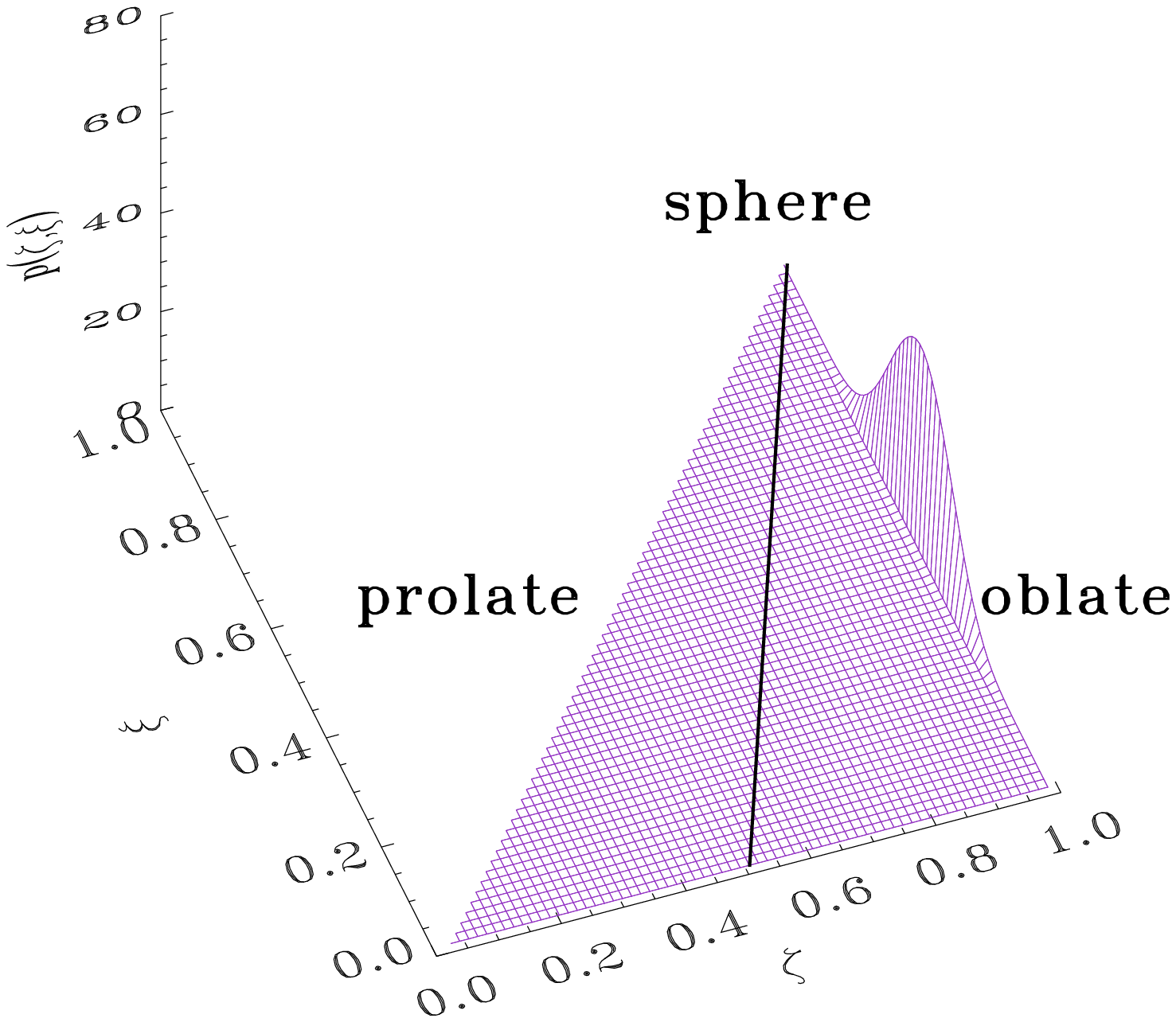}
\caption{Upper panel: maximum-likelihood modified beta-distribution;
  lower panel: maximum-likelihood modified lognormal 
distribution 
}\label{fig1}
\end{figure}

\begin{figure}
\includegraphics[width=2.8in]{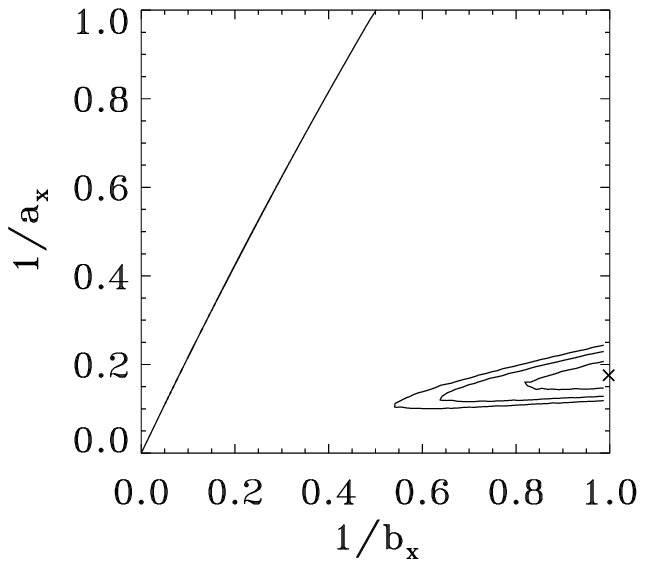}
\includegraphics[width=2.8in]{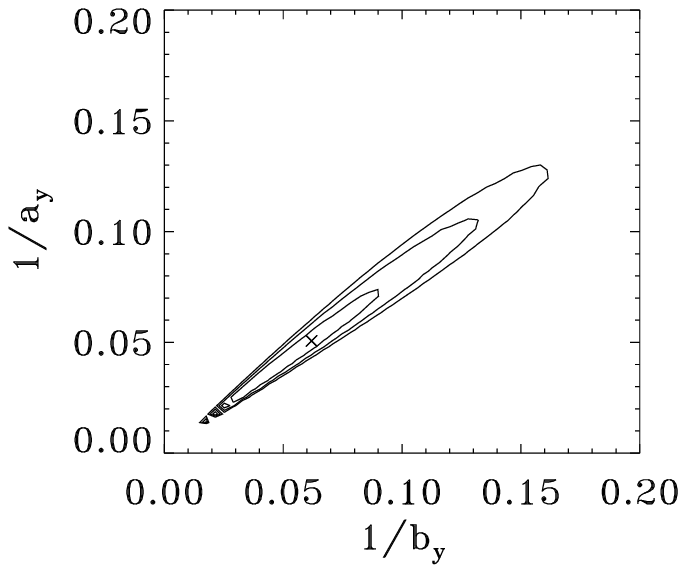}
\caption{Contours of the likelihood function at a level of $\sim 1/3$,
  $\sim 1/22$ and $\sim 1/100$ smaller than the value at the
  maximum. Upper panel: $a_y,b_y$ are fixed at their
  maximum-likelihood values; lower panel: $a_x, b_x$ are fixed at
  their maximum-likelihood values.  }\label{fig2}
\end{figure}

\begin{figure}
\includegraphics[width=2.8in]{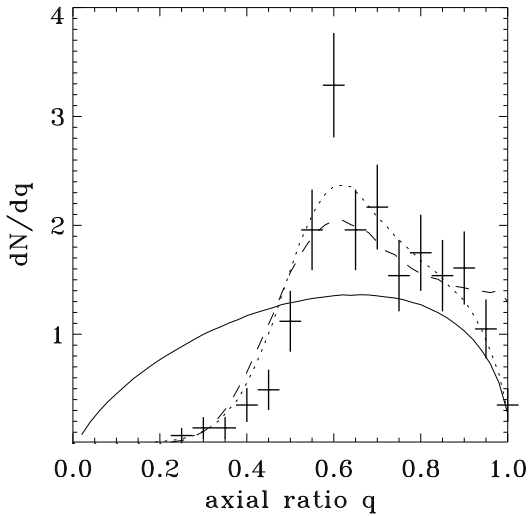}
\caption{Distribution of axial ratios (minor/major axis) 
of projected starless core ellipses
  (datapoints). The lines show the projected core ellipse
  distributions derived from a uniform intrinsic shape distribution
  (solid line), the maximum-likelihood beta-distribution (dotted
  line) and the maximum-likelihood lognormal (dashed line).
}\label{fig3}
\end{figure}

The upper panel of Fig. \ref{fig1} shows surface plots of 
the maximum-likelihood distribution for the
beta-family of distributions. 
Its shape parameters were obtained by maximizing the
likelihood function, using the ``Simulated Annealing'' algorithm (Corana et
al. 1987).
 On the $\zeta-\xi$ plane we have indicated the dividing
line between oblate and prolate cores (oblate cores have $\zeta >
0.5(1+\xi)$).
On this plane, spheres live at $\zeta = \xi = 1$, infinitesimally thin
axisymmetric disks have ($\zeta=1$, $\xi = 0$) and infinitesimally
thin cigars have ($\zeta = 0$, $\xi=0$). Finite-thickness
axisymmetric disks live along the $\zeta = 1$ line. The most likely
distribution of intrinsic shapes is peaked at  $\zeta \approx 1$ and
$\xi \approx 0.54$, indicating that the most probable shape is a
finite-thickness disk with small deviations from
axisymmetry. However, larger degrees of triaxiality are also
frequently encountered, as the distribution decreases only gradually
with decreasing $\zeta$. Even so, only a very small fraction of objects
in this distribution are prolate. The distribution
peaks strongly at oblate shapes. 

To test the sensitivity of our likelihood
analysis to the assumed functional form of the intrinsic shape
distribution, in the lower
panel of Fig. \ref{fig1} we plot the maximum-likelihood
modified lognormal distribution. Although the detailed shape of the two
maximum-likelihood distributions is different, their overall features  
are in remarkable agreement. The distributions
are strongly peaked at about the same point ($\zeta \sim 1$, $\xi\sim
0.5$), indicating that the most
probable intrinsic shape of cores found in the Orion star-forming
regions is predominantly oblate, and the half-thickness of the cores is
finite (about one-quarter the diameter of the disk).   

Qualitatively this result is
to be expected, as most of the elongation measurements indicate
projected
ellipses with low elongations, while there are also a few cores with
significant elongations. 
If viewed from various random angles, an oblate core will most of the time
yield an only mildly elongated projection (face-on or almost face-on
observations), while in the few cases when the core is viewed edge-on
or close to edge-on, a significantly elongated projection will result. In
contrast, a prolate core viewed at random angles will
most of the time yield a significantly elongated projection, a result
not corroborated by the data (see Fig. \ref{fig3} and discussion below).

Figure \ref{fig2} gives a sense of the uncertainty
associated with our determination of the intrinsic shape distribution 
parameters. It shows contours of the likelihood function for the
beta-family of distributions. In the upper panel, we plot contours of
the likelihood as a function of $1/a_x$ and $1/b_x$, with $a_y$ and $b_y$
fixed at their maximum-likelihood values.
The ($a_x$, $b_x$) parameters primarily control the
shape of the distribution along the $\zeta$ axis. 
The straight line represents the division between the regime
where the peak of $p(\zeta,\xi)$ corresponds to an oblate object
(lower right), and the regime where the peak corresponds to a prolate
object (upper left), assuming that the other two parameters
($a_y,b_y$) are fixed at their maximum-likelihood values.  The
contours correspond to a reduction of the level of the likelihood with
respect to its maximum value of $\sim 1/3$,
  $\sim 1/22$ and $\sim 1/100$, which would be the levels of the $1,
  2$ and $3\sigma$ contours for a 2-parameter likelihood with $\log \mathcal{L}$ falling as a chi-square distribution. 
  The likelihood is very
  strongly peaked at the oblate part of the plane, with the likelihood
  of a prolate peak being infinitesimal. 

The lower panel shows the same-level
  contours for the likelihood as a function of $1/a_y$ and $1/b_y$,
  with $a_x$ and $b_x$ fixed at their maximum-likelihood values. The
  $a_y$, $b_y$ parameters primarily control the shape of the
  distribution along the $\xi$ axis. In this case, there is no
  dividing line since the distribution peak would correspond to an
  oblate object for any pair of ($a_y$, $b_y$) for the given values of
  $a_x,b_x$: the latter fix the peak along the $\zeta$ axis very
  close to $\zeta = 1$, and objects with such high $\zeta$ are always
  oblate independently of the value of $\xi$. The
  likelihood is concentrated close to the diagonal, and has a
  preference for  comparable
  values for $a_y$ and
  $b_y$, indicating that, along the $\xi$-axis, the intrinsic shape
  distribution is roughly symmetric. The slight ringing at the
  lower-left corner of the plot is a numerical effect due to the
  finite resolution in our calculation of the likelihood, and does not
  affect the values of our maximum-likelihood parameters since it
  occurs sufficiently far from the likelihood maximum. 

Figure \ref{fig3} shows the distribution of axial ratios for the
projected core ellipses derived from three intrinsic shape
distributions: the maximum-likelihood beta-distribution (dotted line),
the maximum-likelihood lognormal distribution (dashed line), and a
uniform distribution, $p_{\rm Orion}(\zeta,\xi)=$const. (solid line). The crosses
correspond to the NWT07 data, binned in intervals of
$\Delta q = 0.05$. The uniform intrinsic shape
distribution roughly approximates a distribution of core shapes
expected from turbulence-driven cloud fragmentation and
core formation (e.g. Gammie et al. 2003; Li et al. 2004). That the
maximum-likelihood distributions are a better fit to the data is
immediately obvious. We have also performed a
Kolmogorov-Smirnov test to quantify the discrepancy between the data
and the uniform distribution, and to verify that the
maximum-likelihood distributions are in fact likely
in absolute terms to reproduce the observed dataset. 
We find that the probability that the
observed data come from the uniform intrinsic shape distribution is
smaller than $0.1\%$. The same test indicates that the dataset is
consistent within the $1\sigma$ level with originating from either of the maximum-likelihood distributions, which are peaked at
oblate intrinsic shapes.

A potential source of bias in this
analysis is the resolution limitations of the data. It is conceivable
that, for smaller-sized cores, the actual semi-major and semi-minor
axes of the projected core ellipse are both comparable to, or smaller
than, the smallest angle that can be resolved in the data. Such a case
would lead to an observational bias toward circular core ellipses for
the smallest objects, in which case our statistical analysis
 would lead to erroneous results. We have
checked that this is not the case in our dataset. The largest number
of our smallest elongation cores were {\em not}, in fact, close to the
resolution limit. In addition, as we can see in Fig. \ref{fig3}, there is
no ``pileup'' of cores towards $q=1$, as would be expected if the bias
discussed here were significant. 

\section{Discussion}\label{disc}

We have derived the maximum-likelihood beta-distribution of intrinsic
shapes for the Orion GMC starless molecular cloud cores of the NWT07
set. The maximum-likelihood
distribution peaks at the shape of an oblate, finite-thickness, nearly
axisymmetric disk. Oblate cores of varying thickness and varying
degrees of triaxiality also occur frequently, however the
distribution falls very quickly to zero in the prolate part of the
spectrum of shapes. Cores in Orion are most likely intrinsically
oblate. This main result is robust with respect to differences in the
assumed functional form of the intrinsic distribution.
Our result is consistent with the
findings of Jones et al. (2001) and Jones \& Basu (2002),  who used 
different datasets and a different statistical treatment. Hence, it
appears that the result that molecular cloud cores are preferentially
oblate is also robust to the details of the dataset used, the
wavelength of the observations, and the details of the statistical
treatment. A uniform distribution of shapes with equal numbers of oblate and
prolate cores is rejected by the data at a high confidence level ($< 0.1\%$). 

{\em These results are particularly interesting since they can have
important implications for theories of core
formation}. Turbulence-induced core formation tends to favor uniform triaxial shape distributions with, if anything, a slight preference for prolate starless cores (e.g. Gammie et al. 2003; Li et al. 2004). Such distributions are not preferred by the observational data.
In contrast, magnetically-driven fragmentation
naturally leads to oblate objects due to the extra magnetic support
perpendicular to the field lines. 

An additional effect beyond the core formation process which could, in principle, alter the shape of molecular cloud cores and introduce a bias favoring oblate shapes is rotation. However, the ratios of rotational and gravitational energy in prestellar cores such as the ones examined here are observed to be very small (typically $<$ few percent) and there are no observational indications of rotational flattening (e.g. Goodman et al. 1993; Caselli et al. 2002). Even these small amounts of angular momentum can lead to appreciable flattening and rotationally supported discs after sufficient contraction under angular momentum conservation - however, 
such flattening becomes important in much later contraction stages, and at small ($< 10^3 {\rm \, AU}$) scales (overall, protostellar core shapes are not significantly different than those of prestellar cores, e.g. Goodwin et al. 2002). Similar results on the absence of rotationally-induced flattening are also found in simulations (Gammie et al 2003; Li et al. 2004).

The maximum-likelihood approach used in this work is a powerful
statistical technique, which can be appropriately expanded to
explicitly account for observational uncertainties and be combined
with magnetic field orientations from polarization data as those
become available.

Other than observational uncertainties, our results may be biased by
three additional factors. First, in order to derive shapes from the
signal-to-noise submillimeter maps, one makes the implicit assumption
of core isothermality, so that emissivity contours truly correspond
to the spatial distribution of core mass. This source of uncertainty
is not of great concern, however, because on the one hand
isothermality is likely to be an excellent approximation for starless
cores especially at the outer edges of the core defining its shape. On
the other hand, deviations from isothermality would only affect the
shape determination if the temperature gradient does not follow the
density gradient, which is not expected for gravitational heating. 
A second potential
complication is that these cores all originate in the same GMC. If
there is a preferential direction in this cloud
set by the large-scale magnetic field {\em and} if the magnetic field
is dynamically important, then cores may be preferentially oriented
with their smallest ellipsoid axes aligned with the magnetic field,
which might weaken our assumption of random viewing angle
orientations. However, in the turbulent core formation scenario, this
is not a concern since the magnetic field is in this case dynamically
unimportant and tangled. Therefore, the comparison of the observed core distribution at least with the predictions of the turbulence theory of core formation is not affected by such a bias, and this effect cannot ameliorate the discrepancy between roughly uniform shape distributions and the observed data.
Additionally, the survey area from which our dataset was derived spans
a $10^\circ$ region in the sky, rather than a single small cloud where
orientation biases would be expected to be most significant. 
Finally, it is conceivable that the intrinsic shape
distribution is not singly-peaked, but involves two distinct peaks
in different regions of the $(\zeta,\xi)$ parameter space.

\section*{Acknowledgments}
{I thank S. Basu, and V. Pavlidou  for enlightening discussions and 
T. Ch. Mouschovias, A. K\"{o}nigl, M. Kunz, L. Looney, 
and the referee, R. Klessen, for comments 
on the manuscript which improved this paper. 
This work was supported by NSF grants AST 02-06216 and
AST02-39759,  by the NASA Theoretical Astrophysics Program grant
NNG04G178G and the Kavli Institute for Cosmological Physics 
through the grant NSF PHY-0114422.}


\begin{thebibliography}{99}

\bibitem[Basu \& Ciolek(2004)]{2004ApJ...607L..39B} Basu, S., \& Ciolek, 
G.~E.\ 2004, ApJ, 607, L39 

\bibitem[Binney(1985)]{1985MNRAS.212..767B} Binney, J.\ 1985, MNRAS, 212, 
767 

\bibitem[Caselli et al.(2002)]{2002ApJ...572..238C} Caselli, P., Benson,  P.~J., Myers, P.~C., Tafalla, M.\ 2002, ApJ, 572, 238 


\bibitem[Ciolek \& Basu(2006)]{2006ApJ...652..442C} Ciolek, G.~E., \& Basu, 
  S.\ 2006, ApJ, 652, 442 


\bibitem[Corana et al.(1987)]{cor87} Corana, A., Marchesi, M.,
  Martini, C., Ridella, S. 1987, ACM Trans. Math. Soft., 13, 262

\bibitem[Gammie et al.(2003)]{2003ApJ...592..203G} Gammie, C.~F., Lin, 
Y.-T., Stone, J.~M., Ostriker, E.~C.\ 2003, ApJ, 592, 203 

\bibitem[Goodman et al.(1993)]{1993ApJ...406..528G} Goodman, A.~A., Benson, P.~J., Fuller, G.~A., Myers, P.~C.\ 1993, ApJ, 406, 528 

\bibitem[Goodwin et al.(2002)]{2002MNRAS.330..769G} Goodwin, S.~P., 
Ward-Thompson, D., Whitworth, A.~P.\ 2002, MNRAS, 330, 769 

\bibitem[Jones et al.(2001)]{2001ApJ...551..387J} Jones, C.~E., Basu,
  S.,  Dubinski, J.\ 2001, ApJ, 551, 387 

\bibitem[Jones \& Basu(2002)]{2002ApJ...569..280J} Jones, C.~E., Basu, 
S.\ 2002, ApJ, 569, 280 

\bibitem[Kerton et al.(2003)]{2003A&A...411..149K} Kerton, C.~R., Brunt, 
C.~M., Jones, C.~E., Basu, S.\ 2003, A\&A, 411, 149 

\bibitem[Li et al.(2004)]{2004ApJ...605..800L} Li, P.~S., Norman, M.~L., 
Mac Low, M.-M., Heitsch, F.\ 2004, ApJ, 605, 800

\bibitem[Mac Low \& Klessen(2004)]{2004RvMP...76..125M} Mac Low, M.-M., \& 
Klessen, R.~S.\ 2004, Reviews of Modern Physics, 76, 125 

\bibitem[Mouschovias(1976)]{1976ApJ...207..141M} Mouschovias, T.~Ch.\ 1976, 
ApJ, 207, 141 

\bibitem[Myers et al.(1991)]{1991ApJ...376..561M} Myers, P.~C., Fuller, 
G.~A., Goodman, A.~A., Benson, P.~J.\ 1991, ApJ, 376, 561 

\bibitem[Nutter \& Ward-Thompson(2007)]{2007MNRAS.374.1413N} Nutter, D.,  
Ward-Thompson, D.\ 2007, MNRAS, 374, 1413 

\bibitem[Ryden(1996)]{1996ApJ...471..822R} Ryden, B.~S.\ 1996, ApJ, 471, 
822 

\end{thebibliography}
\end{document}